\documentclass[prl,showpacs,amsfonts,amssymb,amsmath,floats,twocolumn,aps]{revtex4}

\usepackage[pdftex]{graphicx}
\usepackage{dcolumn}
\usepackage{bm}
\usepackage{color}

\usepackage[pdftex,colorlinks=true,linkcolor=blue,citecolor=blue,filecolor=blue]{hyperref}

\newcommand{\ff}[1]{{\boldsymbol #1}}
\newcommand{\ca}[1]{{\cal #1}}

\newcommand{\bi}{\begin{itemize}}
\newcommand{\ei}{\end{itemize}}
\newcommand{\be}{\begin{equation}}
\newcommand{\ee}{\end{equation}}

\newcommand{\parag}[1]{\paragraph{#1}}

\begin{document} 

\title{
Relaxation of a classical spin coupled to a strongly correlated electron system
}

\author{Mohammad Sayad, Roman Rausch and Michael Potthoff}
\affiliation{I. Institute for Theoretical Physics, University of Hamburg, Jungiusstra\ss{}e 9, D-20355 Hamburg, Germany}

\begin{abstract}
A classical spin which is antiferromagnetically coupled to a system of strongly correlated conduction electrons 
is shown to exhibit unconventional real-time dynamics which cannot be described by Gilbert damping.
Depending on the strength of the local Coulomb interaction $U$, 
the two main electronic dissipation channels, 
transport of excitations via correlated hopping 
and via excitations of correlation-induced magnetic moments, 
become active on largely different time scales.
We demonstrate that correlations can lead to a strongly suppressed relaxation which so far has been observed in purely electronic systems only 
and which is governed here by proximity to the divergent magnetic time scale in the infinite-$U$ limit.
\end{abstract} 
 


\maketitle 
  
\parag{Motivation.}
A classical spin in an external magnetic field 
shows a precessional motion but when exchange-coupled to a conduction-electron system
the spin additionally relaxes and finally aligns to the field direction. 
This is successfully described on a phenomenological level by the Landau-Lifschitz-Gilbert (LLG) equation 
\cite{llg} and extensions of this concept \cite{TKS08,BMS09}. 
The Gilbert damping constant $\alpha$ is often taken as a phenomenological parameter but can also be computed {\em ab initio} for real materials \cite{GIS07,EMKK11,Sak12} within a framework of effectively independent electrons using band theory \cite{FI11} and then serves as an important input for atomistic spin-dynamics calculations \cite{SHNE08}. 

Electron correlations are expected to have an important effect on the spin dynamics. 
This has been demonstrated in a few pioneering studies \cite{HVT08,GMcD09,HFR15} -- within different models and using various approximations -- but only indirectly by computing the effect of the Coulomb interaction on the Gilbert damping.
One hallmark of strong correlations, however, is the emergence and the separation of energy (and time) scales -- with the correlation-induced Mott insulator \cite{Geb97} as a paradigmatic example. 

With the present study we address correlation effects beyond an LLG-type approach and keep the full temporal memory effect.
It is demonstrated that correlation-induced time-scale separation has profound and qualitatively new consequences for the spin dynamics. 
These are important, e.g., 
for the microscopic understanding of the emerging relaxation time scales in modern nano-spintronics devices involving various transition metals and compounds \cite{Mor10,LEL+10,KWCW11}.

Concretely, we consider a generic model with a classical spin $\ff S$ that is antiferromagnetically exchange coupled ($J>0$) to a Hubbard system and study the spin dynamics as a function of the Hubbard-$U$. 
To tackle this quantum-classical hybrid problem, we develop a novel combination of linear-response theory \cite{Sak12,BNF12,SP15} for the spin dynamics with time-dependent density-matrix renormalization group (t-DMRG) \cite{Sch11,HCO+11,HLO+14} for the correlated electron system.
For technical reasons we consider a Hubbard chain but concentrate on generic effects which are not bound to the one-dimensionality of the model.

In the metallic phase at quarter filling, a complex phenomenology is found where two different channels for energy and spin dissipation, namely dissipation via correlated hopping and via excitations of local magnetic moments, become active on characteristic time scales, depending on $U$. 
While magnetic excitations give the by far dominating contribution to the Gilbert damping in the strong-coupling limit, they contribute to the spin dynamics to a much lesser extent and on later and later time scales when $U$ is increased.

It is demonstrated that electron correlations can have extreme consequences: 
At half-filling and strong $U$, the spin relaxation is incomplete on intermediate time scales. 
This represents a novel effect in a quantum-classical hybrid model which is reminiscent of prethermalization \cite{MK08,MK10,KWE11,MMGS13} or metastability of excitations due to lack of phase space for decay 
\cite{RRBV08,SGJ+10,HP12,RP16}, i.e., physics which so far has been observed in purely electronic quantum systems only.

\parag{Gilbert damping.}
We consider the Hubbard model for $N$ electrons on an open chain of length $L$ as a prototypical model of correlated conduction electrons:
\be
  H_{\rm e} = - T \sum_{i<j}^{n.n.} \sum_{\sigma} (c_{i\sigma}^{\dagger} c_{j\sigma} + \mbox{H.c.})
  + U \sum_{i=1}^{L} n_{i\uparrow} n_{i\downarrow} \: .
\label{eq:he}  
\ee 
The nearest-neighbor (n.n.) hopping $T=1$ sets the energy and time scale ($\hbar=1$). 
Using standard arguments \cite{BNF12,SP15}, the Gilbert damping parameter $\alpha$ can be computed as
\be
\alpha = - J^{2} \int_{0}^{\infty} dt \, t \, \chi_{\rm loc}(t) 
\label{eq:alpha}  
\ee
and depends on the Hubbard interaction $U$ via the local (diagonal and isotropic) retarded spin susceptibility 
\be
  \chi_{\rm loc}(t) = - i \Theta(t) \langle 0 | [s_{i_{0}z}(t) , s_{i_{0}z}(0)] | 0 \rangle
\label{eq:chi}  
\ee
at the site $i_{0}$ where the classical spin is coupled to. 
$J$ is the strength of the exchange interaction [see Eq.\ (\ref{eq:sp}) below].
Furthermore, $|0\rangle$ is the ground state of $H_{\rm e}$, 
$s_{iz}(t) = e^{iH_{\rm e}t} s_{iz} e^{-iH_{\rm e}t}$, and 
$s_{iz}$ is the $z$-component of the local conduction-electron spin $\ff s_{i} = \sum_{\sigma\sigma'} c^{\dagger}_{i\sigma} \ff \tau_{\sigma\sigma'} c_{i\sigma'} / 2$, with the vector of Pauli matrices $\ff \tau$ and with $\sigma,\sigma' = \uparrow,\downarrow$.

We choose $i_{0}=1$ for two reasons: 
(i) As compared to the symmetric choice $i_{0}=L/2$, this allows us to double the accessible time scale (before finite-size effects set in).
(ii) The time-integral in (\ref{eq:alpha}) is sensitive to the long-time behavior of $\chi_{\rm loc}(t)$ which, at least for $U=0$, is related to the strength of the van Hove singularities in the local density of states \cite{SP15}. 
At the edge of the open chain, those are weak and characteristic for a three-dimensional system.

\parag{Local spin correlations at quarter filling.}
To compute $\chi_{\rm loc}(t)$, we apply t-DMRG and the framework of matrix-product states \cite{Sch11} for systems with $L=80$--$120$ sites. 
Concretely, we use the two-site version of the algorithm suggested in Ref.\ \cite{HCO+11,HLO+14} which is based on the time-dependent variational principle (TDVP).

\begin{figure}[t]
\centering
\includegraphics[width=0.95\columnwidth]{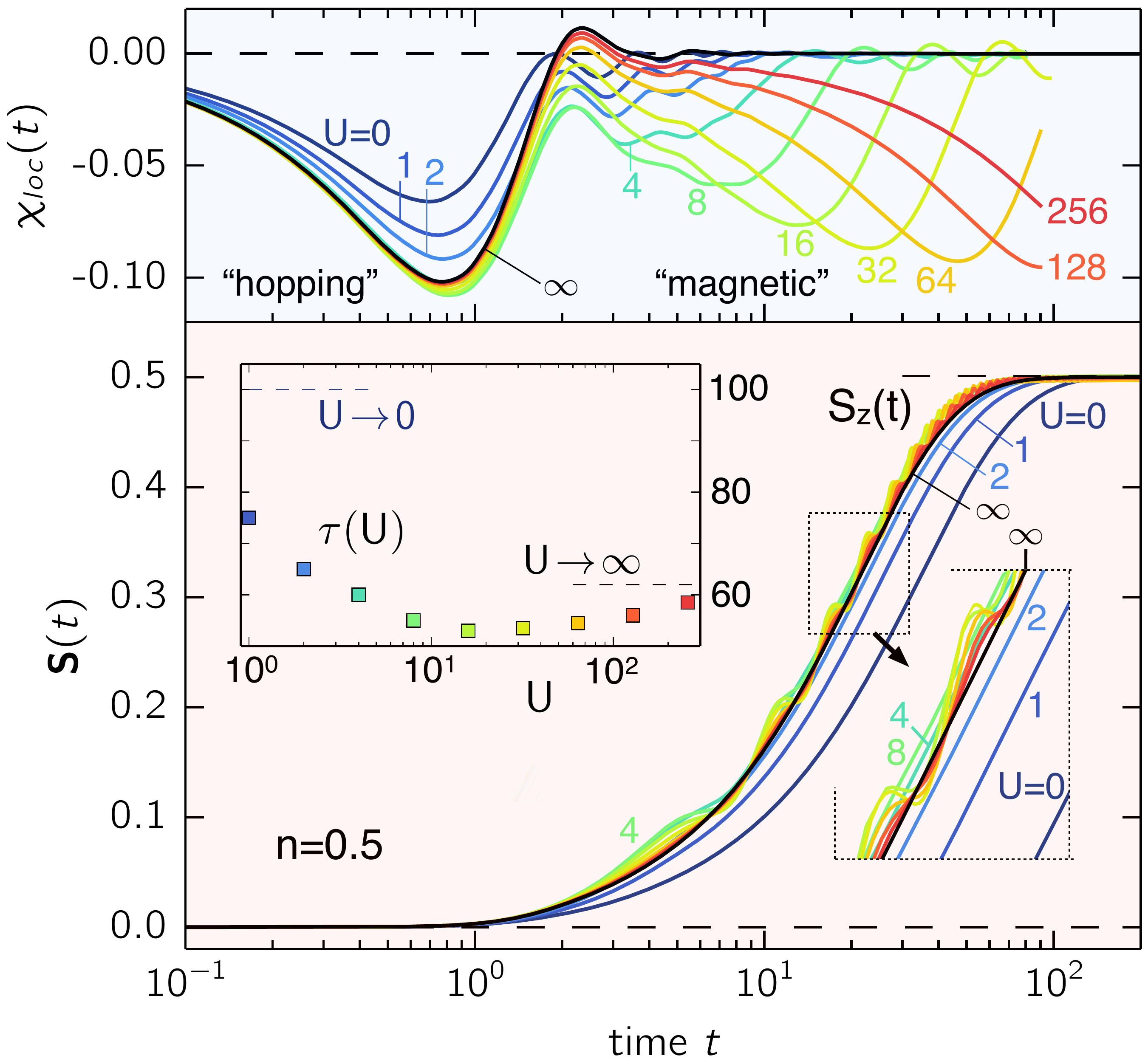}
\caption{(Color online)
{\em Upper panel:}
Local spin correlation $\chi_{\rm loc}(t)$ at $i_{0}=1$ for an open Hubbard chain with $L=80$ sites as obtained by t-DMRG for quarter filling and different $U$ as indicated.
($U\ge 32$: $t-J$ model with three-site terms \cite{ATT95}, $L=100$ sites; $U=\infty$: $L=120$).
Energy and time scales are fixed by the n.n.\ hopping $T=1$.
{\em Lower panel:}
Resulting real-time dynamics of a classical spin $\ff S(t)$ (with $|\ff S(t)|=\frac12$, only $S_{\rm z}$ is shown) coupled at $i_{0}$ to the local conduction-electron spin as obtained from Eq.\ (\ref{eq:id}) for $J=1$ and different $U$.
The spin dynamics is initiated by switching the local magnetic field in Eq.\ (\ref{eq:id}) at time $t=0$ from $x$- to $z$-direction $(B_{\rm fin} = 1)$. 
{\em Inset:} $U$-dependence of the relaxation time $\tau$, defined as $S_{\rm z}(\tau)=0.98 |\ff S|$.
} 
\label{fig:n05}
\end{figure}

Electron correlations are expected to speed up the relaxation of the classical spin since electron scattering facilitates the transport of energy and spin density from $i_{0}$ to the bulk of the system.
An increasingly efficient dissipation implies an increase of $\alpha$ with $U$.
This can be nicely seen in $\chi_{\rm loc}(t)$, which determines $\alpha$ via Eq.\ (\ref{eq:alpha}) and which is shown in Fig.\ \ref{fig:n05} (upper panel) for quarter filling $n \equiv N/L = 0.5$ where we have a (correlated) metal in the entire $U$ range.
In fact, the absolute value of the integral weight $\int dt \, \chi_{\rm loc}(t)$ grows with increasing $U$. 

Note that via the fluctuation-dissipation theorem, the total weight $\int dt \, \chi_{\rm loc}(t)$ is given by the negative local static spin susceptibility. 
This explains that $\chi_{\rm loc}(t)$ is mainly negative.

\parag{Separation of time scales.}
A central observation is that $\chi_{\rm loc}(t)$ develops a pronounced two-peak structure for strong $U$.
For $U=0$ and in the weak-coupling regime, there is essentially a single (negative) peak around $t \sim 1$ only.
This corresponds to fast correlated-hopping processes on a scale set by the inverse hopping $1/T$. 
As is seen in the figure, the contribution of these processes to the Gilbert damping grows with increasing $U$.

The second (negative) peak is clearly present for $U \gtrsim 8$. 
The almost linear shift of its position with $U$ hints towards a time scale set by an effective magnetic interaction $J_{\rm H} \sim 1/U$ between local magnetic moments formed by strong correlations in the conduction-electron system.
Even for $U\to \infty$, however, local-moment formation is not perfect at quarter filling: 
We have $\langle \ff s_{i}^{2} \rangle = \frac34 n = \frac38 < s(s+1)$ with $s=1/2$ for the size of the correlated local moment \cite{Geb97}.
This explains the residual contributions from correlated hopping processes (first peak). 

For strong $U$ the Gilbert damping is dominated by magnetic processes: 
Because of the extra factor $t$ under the integral in Eq.\ (\ref{eq:alpha}), the contribution of the second peak in $\chi_{\rm loc}(t)$ by far exceeds the hopping contribution (note the logarithmic scale in Fig.\ \ref{fig:n05}). 
Clearly, $\alpha$ strongly increases with $U$ though a precise value cannot be given due to limitations of the t-DMRG in accessing the long-time limit.

\parag{Spin-dynamics model.}
The simple LLG equation for a classical spin, i.e., $\dot{\ff S} = \ff S \times \ff B - \alpha \ff S \times \dot{\ff S}$, can only provide an overall picture of the spin dynamics and in fact ignores the electronic time-scale separation. 
We therefore apply a refined approach which explicitly accounts for the conduction-electron degrees of freedom in a model $H$ which, besides $H_{\rm e}$, Eq.\ (\ref{eq:he}), includes the local and isotropic coupling between $\ff s_{i_{0}}$ and the classical spin $\ff S$ ($|\ff S| = 1/2$): 
\begin{equation}
H 
= 
H_{\rm e} + H_{\rm e-spin} 
=
H_{\rm e} 
+
J \ff s_{i_0} \ff S 
-
\ff B \ff S \; .
\label{eq:sp}
\end{equation}
We have also added a local magnetic field $\ff B$ which, at time $t=0$, is suddenly switched from $\ff B = B_{\rm ini} \hat{x}$, forcing the spin to point in $x$ direction, 
to $\ff B = B_{\rm fin} \hat{z}$ with $B_{\rm fin}=1$ to initiate the spin dynamics.
Complete relaxation is achieved if $\ff S(t) \to \frac12 \hat{z}$ for $t\to \infty$.

The Hamiltonian (\ref{eq:sp}) represents a semiclassical Kondo-impurity model with finite Hubbard-$U$. 
For a strong local field $B_{\rm fin}$, the dynamical Kondo effect \cite{NGA+15} that would show up in case of a quantum-spin $S=\frac12$ is suppressed -- this justifies the classical-spin approximation. 
The quantum-classical hybrid (\ref{eq:sp}) also results in the limit of large spin quantum numbers $S$ \cite{Gar08} of a correlated quantum-spin Kondo impurity model \cite{SRP16}.

$\ff S(t)$ satisfies the classical equation of motion $\dot{\ff S}(t) = \ff S(t) \times \ff B - J \ff S(t) \times \langle \ff s_{i_0} \rangle_{t}$ \cite{Elz12}.
Applying lowest-order perturbation theory in $J$, the Kubo formula yields $\langle \ff s_{i_{0}} \rangle_{t} =  J \int_{0}^{t} dt' \, {\chi_{\rm loc}}(t-t') \ff S(t')$ where the retarded local spin correlation $\chi_{\rm loc}(t)$ now plays the role of the linear-response function.
As has been demonstrated in Ref.\ \cite{SP15} for $U=0$, this approach is perfectly reliable even for fairly strong couplings $J$ and up to the time scale necessary for complete spin relaxation.  
Here, we choose $J=1$ to generate relaxation times accessible to the t-DMRG approach. 

The resulting effective integro-differential equation of motion \cite{Sak12,BNF12} (see also Refs.\ \cite{retard1,retard2}), 
\begin{equation}
\dot{\ff S}(t)
= 
\ff S(t) \times \ff B \\
- 
J^{2} \ff S(t) \times \int_{0}^{t} dt' \, \chi_{\rm loc}(t-t') \ff S(t') \; , 
\label{eq:id}   
\end{equation}
is numerically solved using a high-order Runge-Kutta \cite{Ver09} and quadrature techniques.

\parag{Correlation effects in the spin dynamics.}
The resulting spin dynamics (Fig.\ \ref{fig:n05}, lower panel) is characterized by precessional motion ($S_{\rm x}(t), S_{\rm y}(t)$ not shown) with Larmor frequency $\omega_{L} \propto B$ around the $\hat{z}$ axis, and by relaxation driven by dissipation of energy and spin into the bulk of the electronic system.
In the final state there is complete alignment, $\ff S(t) \uparrow\uparrow \ff B$.

Comparing the results for the different $U$, we find that
(i) significant relaxation starts at times $t \sim 1/T$, i.e., on the time scale for dissipation through correlated-hopping processes. 
(ii) Correlation effects lead to a considerably shorter relaxation time, e.g., by about a factor two when comparing the results for $U=0$ and $U=16$ (see inset). 
(iii) To some extent this is due to an additional damping mechanism, namely via excitations of {\em correlation-induced} magnetic moments -- at least for moderate $U$. 
(iv) For strong $U$, however, the relaxation time {\em increases} again. 
This is counterintuitive but easily explained: 
Since the second, ``magnetic'' peak in $\chi_{\rm loc}(t)$ shifts with increasing $U$ to later and later times, relaxation is already completed before dissipation through spin-flip processes can become active.
This is most obvious for $U \to \infty$ where magnetic damping is {\em never} activated, and where a renormalized band picture may apply.
(v) At intermediate $U$, however, the picture is different. 
Here, spin-flip processes do contribute to the relaxation but more than an order of magnitude later ($t \gtrsim U / T^{2}$) than the hopping time scale. 
{\em Despite their dominating contribution to} $\alpha$, their effect is weaker as compared to the correlated-hopping processes. 
Still, spin-flips leave a clear characteristics in $S_{z}(t)$: 
their additional torque produces oscillations (with a period largely independent of $U$) which superimpose the monotonic relaxation dynamics. 
The onset of these ``magnetic'' oscillations linearly grows with $U$. 
Note that the interpretation of these (and the following) findings does not rely on the one-dimensionality of the model.

\begin{figure}[t]
\centering
\includegraphics[width=0.95\columnwidth]{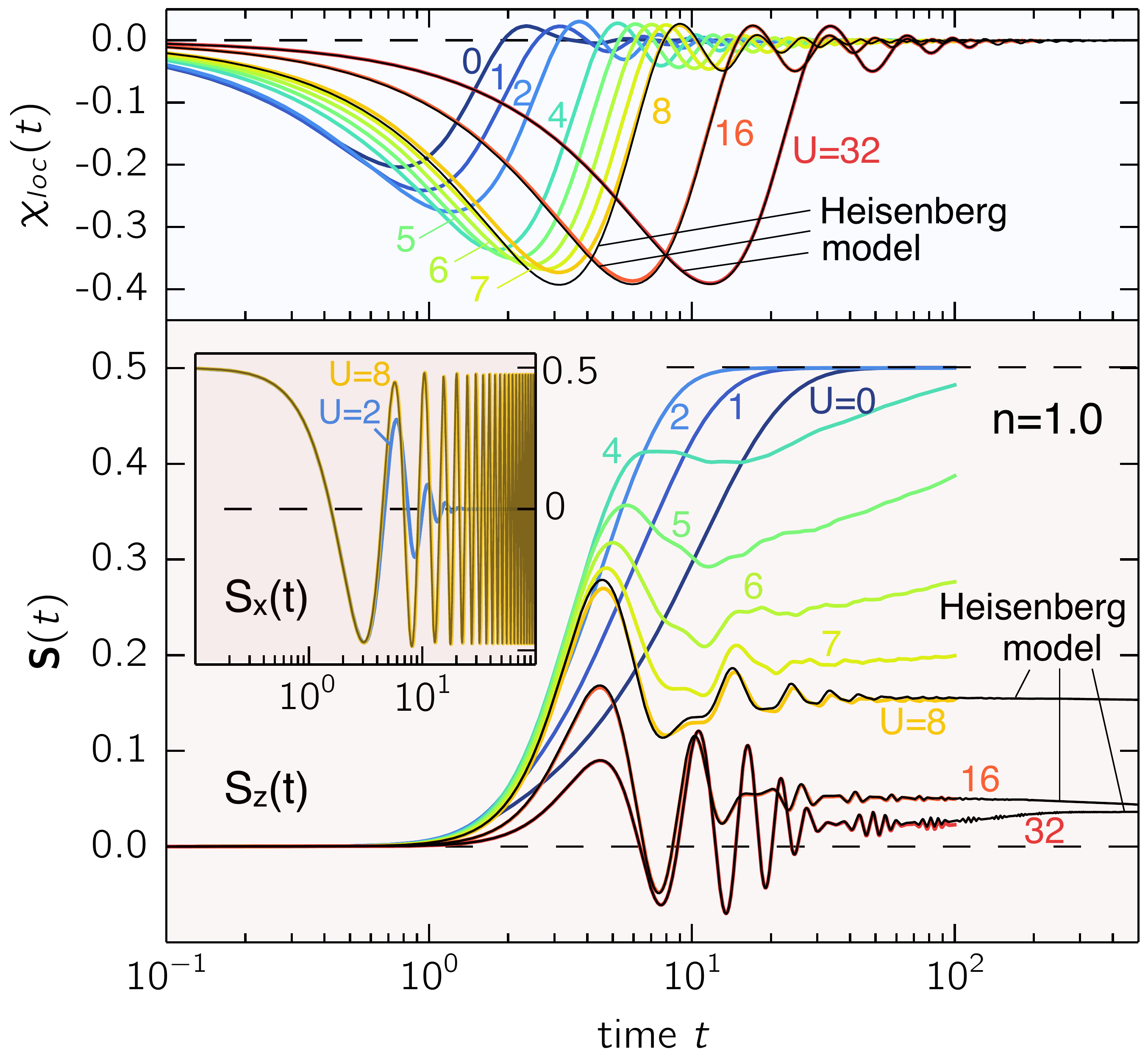}
\caption{(Color online)
The same as Fig.\ \ref{fig:n05} but for $n=1$ ($L=60$).
Thin black lines: 
Heisenberg model with $J_{\rm H} = \frac{4T^{2}}{U}$ ($L=400$) and, for improved accuracy at $U=8$, 
with n.n.\ and n.n.n.\ couplings $J_{\rm H} = \frac{4T^{2}}{U} - \frac{16T^{4}}{U^{3}}$ and $J'_{\rm H} = \frac{4T^{4}}{U^{3}}$ \cite{Bul67} ($L=300$).
} 
\label{fig:n10}
\end{figure}

\parag{Gilbert damping of a Mott insulator.}
Dissipation through correlated hopping is impeded or even suppressed at half-filling where the system is a Mott insulator for all $U>0$. 
Fig.\ \ref{fig:n10} (upper panel) shows the t-DMRG data for $\chi_{\rm loc}(t)$ at $n=1$ and different $U$.
Its time dependence is dominated by a single (negative) structure which grows with increasing $U$ up to, say, $U \approx 8$. 
In the weak-coupling regime, $U\lesssim 4$, the local magnetic moments are not yet well-formed since the charge gap $\Delta \sim e^{-1/U}$ (as obtained from the Bethe ansatz \cite{Ovc69} for $U\to 0$) is small as compared to $T$.
Hence, residual hopping processes still contribute significantly.

In the strong-coupling limit, on the other hand, spin-flip processes dominate.
Here, we observe scaling behavior, $\chi_{\rm loc}(t) = F(4t T^{2}/{U})$ with a universal function $F(x)$. 
Indeed, due to the suppression of charge fluctuations, the long-time, low-energy dynamics is captured by a Heisenberg chain $H_{\rm Heis.} = J_{\rm H} \sum_{i} \ff s_{i} \ff s_{i+1}$ with antiferromagnetic interaction $J_{\rm H} = 4T^{2}/U$ between rigid $s=1/2$-spins. 
As $J_{\rm H}$ is the only energy scale remaining, $F(t J_{\rm H})$ is the retarded local susceptibility of the Heisenberg chain.
With $F(x)$ obtained numerically by means of t-DMRG applied to $H_{\rm Heis.}$ at $J_{\rm H}=1$, the t-DMRG data for strong $U$ are fitted perfectly (see Fig.\ \ref{fig:n10}).
Significant deviations from the scaling behavior can be seen in Fig.\ \ref{fig:n10} for $U = 8$ and $t \approx 3$, for instance.

Scaling can be exploited to determine the $U$-dependence of the Gilbert damping for a Mott insulator. 
From Eq.\ (\ref{eq:alpha}) we get
\be
  \alpha = \frac{J^{2}}{J_{\rm H}^{2}} \int_{0}^{\infty} dx \, x \, F(x) = \frac{J^{2}}{J_{\rm H}^{2}} \alpha_{0}
  = \frac{J^{2} U^{2}}{16 T^{4}} \, \alpha_{0} \: , 
\label{eq:scale}
\ee
and thus, for fixed $J, T$, we have $\alpha \propto U^{2}$.
For the universal dimensionless Gilbert damping constant $\alpha_{0}$ we find
\be
  \alpha_{0} \approx 4.8 \: .
\label{eq:univ}
\ee
For a correlated Mott insulator, Eqs.\ (\ref{eq:scale}) and (\ref{eq:univ}) completely describe the $U$-dependence of the classical-spin dynamics in the weak-$J$, weak-$B$ limit where the $t$-dependence of $\ff S(t)$ is so slow, as compared to the typical memory time $\tau_{\rm mem}$ characterizing $\chi_{\rm loc}(t)$, that the Taylor expansion $\ff S(t') \approx \ff S(t) + \dot{\ff S}(t) (t'-t)$ can be cut at the linear order under the $t'$-integral in Eq.\ (\ref{eq:id}), such that the LLG equation is obtained as a Redfield equation \cite{BF02}.

\parag{Incomplete spin relaxation.}
As demonstated with Fig.\ \ref{fig:n10} (lower panel), there is an anomalous $U$-dependence of the spin dynamics at $n=1$. 
Only in the weak-coupling regime, $U\lesssim 2$, do damping effects increase and lead to a decrease of the relaxation time with increasing $U$.
For $U=4$, however, the relaxation time {\em increases} again. 
This behavior is clearly beyond the LLG theory and is attributed to the fact that 
the memory time, $\tau_{\rm mem} \propto 1/J_{\rm H} \propto U$ for strong $U$, becomes comparable to and finally exceeds the precession time scale $\tau_{\rm B} = 2\pi/B$
(see the Supplementary Material \cite{suppl}).

In addition, as for $n=0.5$, we note a non-monotonic behavior of $S_{z}(t)$ with superimposed oscillations (see $U=6$, for example). 
With increasing $U$ these oscillations die out, and at a ``critical'' interaction $U_{\rm c} \sim 8$ and for all $U>U_{\rm c}$ the relaxation time seems to diverge.
Namely, the $z$-component of $\ff S(t)$ approaches a nearly constant value which decreases with increasing $U$ while $S_{x}$ (and $S_{y}$) still precess around $\ff B$ (see inset). 
Hence, on the accessible time scale, $U_{\rm c}$ marks a transition or crossover to an incompletely relaxed but ``stationary'' state. 

The same type of dynamical transition is also seen for a classical spin coupled to a Heisenberg chain for which much larger system sizes ($L=400$) and thus about an order of magnitude longer time scales are accessible to t-DMRG. 
Here, the crossover coupling is $J_{\rm H, c} \sim 0.5$.
However, these calculations as well as analytical arguments clearly indicate that a state with $S_{z}=\mbox{const.} \ne 1/2$ is unstable and that finally, for $t\to \infty$, the fully relaxed state with $\ff S(t) \uparrow\uparrow \ff B$ is reached
(see \cite{suppl} for details).

The ``stationary state'' on an intermediate time scale originates when the bandwidth of magnetic excitations gets smaller than the field -- as can be studied in detail already for $U=0$ (and very strong $B$).
On the time axis, the missing relaxation results from a strong memory effect which, in the strong-$U$ limit, shows up for $J_{\rm H} \lesssim B$. 
Here, $\tau_{\rm mem} \gtrsim \tau_{B}$ which implies that the $z$-component of the spin torque on $\ff S(t)$ averages to zero \cite{suppl}.

The incomplete spin relaxation can also be understood as a transient ``phase'' similar to the concept of a prethermalized state.
The latter is known for purely electronic systems \cite{MK08,MK10,KWE11,MMGS13} which, in close parametric distance to integrability, do not thermalize directly but are trapped for some time in a prethermalized state. 
Here, for the quantum-classical hybrid, the analogue of an ``integrable'' point is given by the $U\to\infty$ limit where, for every {\em finite} $t$, the integral kernel $\chi_{\rm loc}(t) \equiv 0$, and Eq.\ (\ref{eq:id}) reduces to the simple (linear) Landau-Lifschitz equation \cite{llg}.

The situation is also reminiscent of {\em quantum} excitations which are metastable on an exponentially long time scale due to a small phase space for decay. 
An example is given by doublons in the Hubbard model which, for $U$ much larger than the bandwidth and due to energy conservation, can only decay in a high-order scattering process \cite{RRBV08,SGJ+10,HP12,RP16}.
The relaxation time diverges in the $U\to \infty$ limit where 
the doublon number is conserved.
Here, for a {\em classical} spin, one would expect that relaxation via dissipation of (arbitrarily) small amounts of energy is still possible.
Our results show, however, that this would happen on a longer time scale not accessible to the linear-response approach while the ``stationary state'' on the intermediate time scale is well captured \cite{suppl}.

\parag{Outlook.}
Correlation-induced time-scale separation and incomplete relaxation represent phenomena with further general implications. 
While slow correlation-induced magnetic scales dominate the Gilbert damping $\alpha$, their {\em activation} has been found to depend on microscopic details.
This calls for novel correlated spin-dynamics approaches.
The combination of t-DMRG with non-Markovian classical spin dynamics is an example how to link the fields of strongly correlated electron systems and spin dynamics, but further work is necessary.
Combination with dynamical mean-field theory \cite{dmft} is another promising option. 
Also spin dynamics based on LLG-type approaches combined with {\em ab initio} band theory could be successful in the case of very strong $U$ where due to the absence of magnetic damping a renormalized band picture may be adequate. 
Further progress is even needed for the very theory of a consistent hybrid-system dynamics \cite{Elz12,Sal12}. 
Generally, hybrid systems are not well understood and call for a merger of known quantum and classical concepts, such as eigenstate thermalization, prethermalization, (non-)integrability etc \cite{RDO06}.
However, also concrete practical studies with classical spins coupled to conduction electrons \cite{SP15} are needed, as those hold the key for the microscopic understanding of nano-spintronics devices \cite{KWCW11} or skyrmion dynamics \cite{FCS13,IMN13}.

\parag{Acknowledgements.}
We thank F.\ Hofmann for instructive discussions.
Support of this work by the 
DFG within the SFB 668 (project B3) and within the SFB 925 (project B5) is gratefully acknowledged.

\newpage

\appendix 

\subsection*{\color{blue} Supplementary material}

\subsection{Validity of linear-response theory.}

\begin{figure}[b]
\centering
\includegraphics[width=0.95\columnwidth]{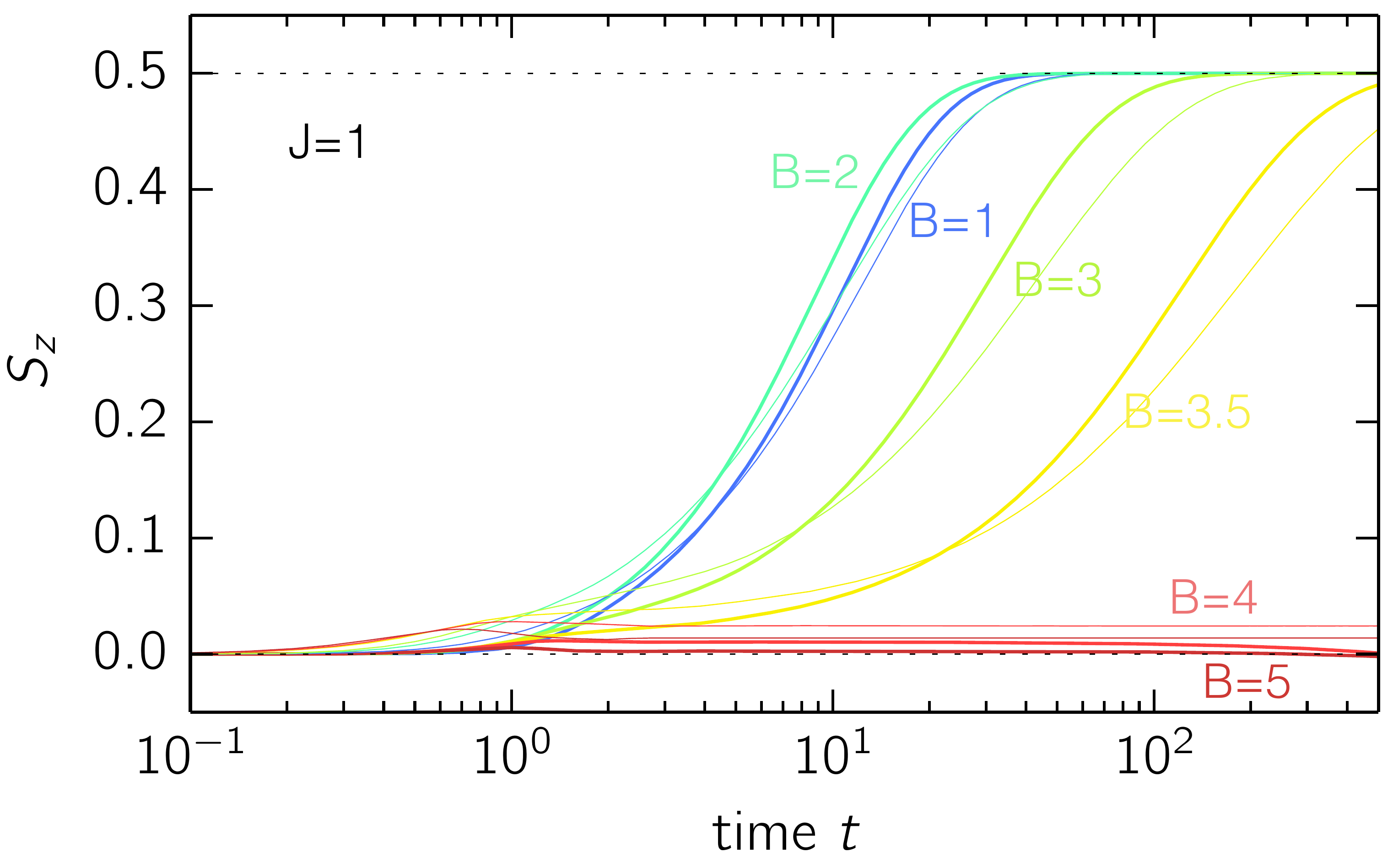}
\caption{
Time dependence of the $z$-component of the classical spin for $J=1$, $n=1$, $U=0$ and different values of the field $B$ as indicated.
Calculations based on the linear-response approach (fat solid lines) are compared to the results of the full quantum-classical hybrid theory (thin solid lines) for $L=500$.
} 
\label{fig:linear}
\end{figure}

The reliability of the linear-response approach (see Eq.\ (5) of the main text) can be tested by comparing with the results of the full (non-perturbative) quantum-classical hybrid dynamics for the model given by Eq.\ (4) of the main text. This is easily accessible for the case $U=0$ (see Ref.\ \cite{SP15} for details).
Fig.\ \ref{fig:linear} displays the time dependence of the $z$-component of the classical spin for $J=1$, for a half-filled system ($n=1$) of non-interacting conduction electrons ($U=0$) and for different strengths of the field $B$ after switching from $x$- to $z$-direction.

While there are some discrepancies visible, as expected, the figure demonstrates that the agreement on a qualitative level is in fact excellent for weak as well as for strong fields. 
Both approaches also predict a crossover from complete to incomplete spin relaxation at $B=B_{\rm c} \approx 4$.
We conclude that the linear-response approach provides reliable results for the classical spin dynamics. 

This can be explained by the observation that $|\langle \ff s_{i_{0}} \rangle_{t}|$ is small and that the classical spin $\ff S(t)$ and the conduction-electron moment $\langle \ff s_{i_{0}} \rangle_{t}$ are nearly collinear at any instant of time (see Ref.\ \cite{SP15} for a detailed and systematic discussion). 
Hence, even for moderately strong couplings $J$, the linear-response contribution $J^{2} \ff S(t) \times \langle \ff s_{i_{0}} \rangle_{t}$ to the equation of motion for $\ff S(t)$ is small (and the quadratic and higher-order corrections are expected to be even smaller).

\subsection{Mechanism for incomplete relaxation.}

Fig.\ \ref{fig:linear} shows that the relaxation of the classical spin becomes incomplete for strong $B$. 
On the basis of the linear-response theory this can be explained as follows: 
The $x$ and $y$ components of the linear response
\begin{equation}
\langle \ff s_{i_{0}} \rangle_{t}
= 
\int_{0}^{t} d\tau \, \chi_{\rm loc}(\tau) \ff S(t-\tau) \; , 
\end{equation}
tend to zero if the characteristic memory time $\tau_{\rm mem}$ of the kernel $\chi_{\rm loc}(\tau)$ is much larger than the precession time scale $\tau_{\rm B} = 2\pi / B$ since the integral produces a vanishing average in this case.
This means that the corresponding torque, $- J^{2} \ff S(t) \times \langle \ff s_{i_{0}} \rangle_{t}$, is perpendicular to the field direction and hence there is no relaxation of the spin.

The same argument can also be formulated after transformation to frequency space:
After some transient effect, we have
$\langle \ff s_{i_{0}} \rangle_{\omega} = \chi_{\rm loc}(\omega) \ff S(\omega)$, 
and thus the $x, y$-components of the linear response will vanish if $\chi_{\rm loc}(\omega=B) = 0$, i.e., if $B$ is stronger than the bandwidth of the magnetic excitations (here: $B_{\rm c} \approx 4$).
Note that this requires an unrealistically strong field in case of non-interacting conduction electrons. 

In the case of correlated conduction electrons, $|\ff S(t) \times \langle \ff s_{i_{0}} \rangle_{t}|$ remains small (of the order of 0.1 or smaller), for weak and for strong $B$, as has been checked numerically.
We therefore expect the linear-response approach to provide qualitatively correct results for $U>0$ as well. 

At half-filling and for strong $U$, the memory time $\tau_{\rm mem} \propto J_{\rm H}^{-1} \propto U$, i.e., $\tau_{\rm mem}$ can easily become large as compared to $\tau_{\rm B}$, and thus incomplete spin relaxation can occur at comparatively weak and physically meaningful field strengths.
For example, from the Bethe ansatz \cite{EFG05} we have
\begin{equation}
  W_{\rm spinon} = 2 \int_{0}^{\infty} \frac{dx}{x} \, \frac{J_{1}(x)}{\cosh(Ux/4)}  \to \frac{\pi}{2} J_{\rm H} \quad \mbox{for} \; U \to \infty
\end{equation}
for the spinon bandwidth $W_{\rm spinon}$ where $J_{1}(x)$ is the first Bessel function.
Hence, for strong Hubbard interaction, $B_{\rm c} \approx 2 W_{\rm spinon} = \pi J_{\rm H}$. 

\subsection{Classical spin coupled to a Heisenberg model.}

At half-filling and in the limit $U\to \infty$ the low-energy physics of the Hubbard model is captured by an antiferromagnetic Heisenberg model, 
\be
  H_{s} = \sum_{i} (J_{\rm H} \ff s_{i} \ff s_{i+1} + J'_{\rm H} \ff s_{i} \ff s_{i+2}) \; , 
\ee
where up to order $\ca O(T^{2}/U)$ the nearest-neighbor and the next-nearest-neighbor couplings \cite{Bul67} are $J_{\rm H} = 4 T^{2}/U$ and $J'_{\rm H} = 0$, and up to order $\ca O(T^{4}/U^{3})$,
\be
J_{\rm H} = \frac{4T^{2}}{U} - \frac{16T^{4}}{U^{3}} \: , \quad J'_{\rm H} = \frac{4T^{4}}{U^{3}} \: .
\label{eq:jh}
\ee
Analytically, by perturbation theory in $x = 4t T^{2} / U = t J_{\rm H}$, one verifies the linear short-time behavior 
\be
\chi_{\rm loc}(t) = \Theta(t) \, t \, \frac23 (J_{\rm H} \langle \ff s_{i_{0}} \ff s_{i_{0}+1} \rangle +
J'_{\rm H} \langle \ff s_{i_{0}} \ff s_{i_{0}+2} \rangle) + \ca O(x^{2}) \: , 
\label{eq:short}
\ee 
valid to leading order for both, the Hubbard and the effective Heisenberg model. 
However, as can be seen in Fig.\ 2 of the main text, the Heisenberg dynamics also applies to intermediate times; the effective model with coupling constants (\ref{eq:jh}) almost perfectly reproduces the results of the Hubbard model for $U \ge 8$. 

\begin{figure}[t]
\centering
\includegraphics[width=0.85\columnwidth]{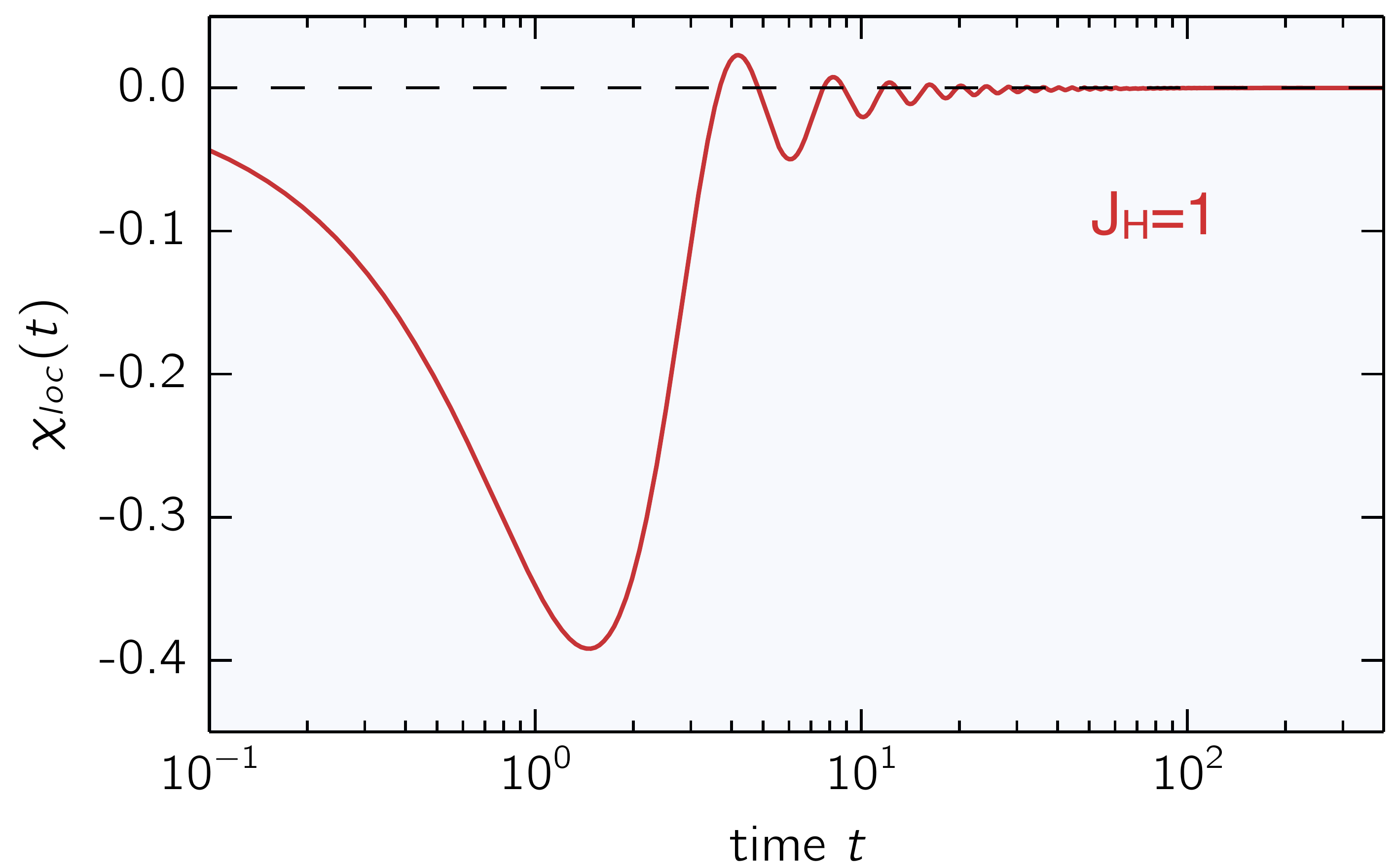}
\caption{
Local susceptibility at the edge of an open Heisenberg chain ($J_{\rm H}=1$). 
} 
\label{fig:heichi}
\end{figure}

Here, we treat the Heisenberg model with n.n.\ coupling as an independent system. 
Fig.\ \ref{fig:heichi} shows the corresponding local spin susceptibility for $J_{\rm H}=1$ as obtained by t-DMRG calculations with $L=400$ Heisenberg spins. 
Since $J_{\rm H}$ is the only energy scale, we have $\chi_{\rm loc}(t) = F(t \, J_{\rm H})$ for arbitrary $J_{\rm H}$ where $F(x)$ is a function independent of $J_{\rm H}$. 
This implies that the dominant (negative) peak of $\chi_{\rm loc}(t)$ shifts to later and later times as $J_{\rm H}$ decreases.

\begin{figure}[t]
\centering
\includegraphics[width=0.95\columnwidth]{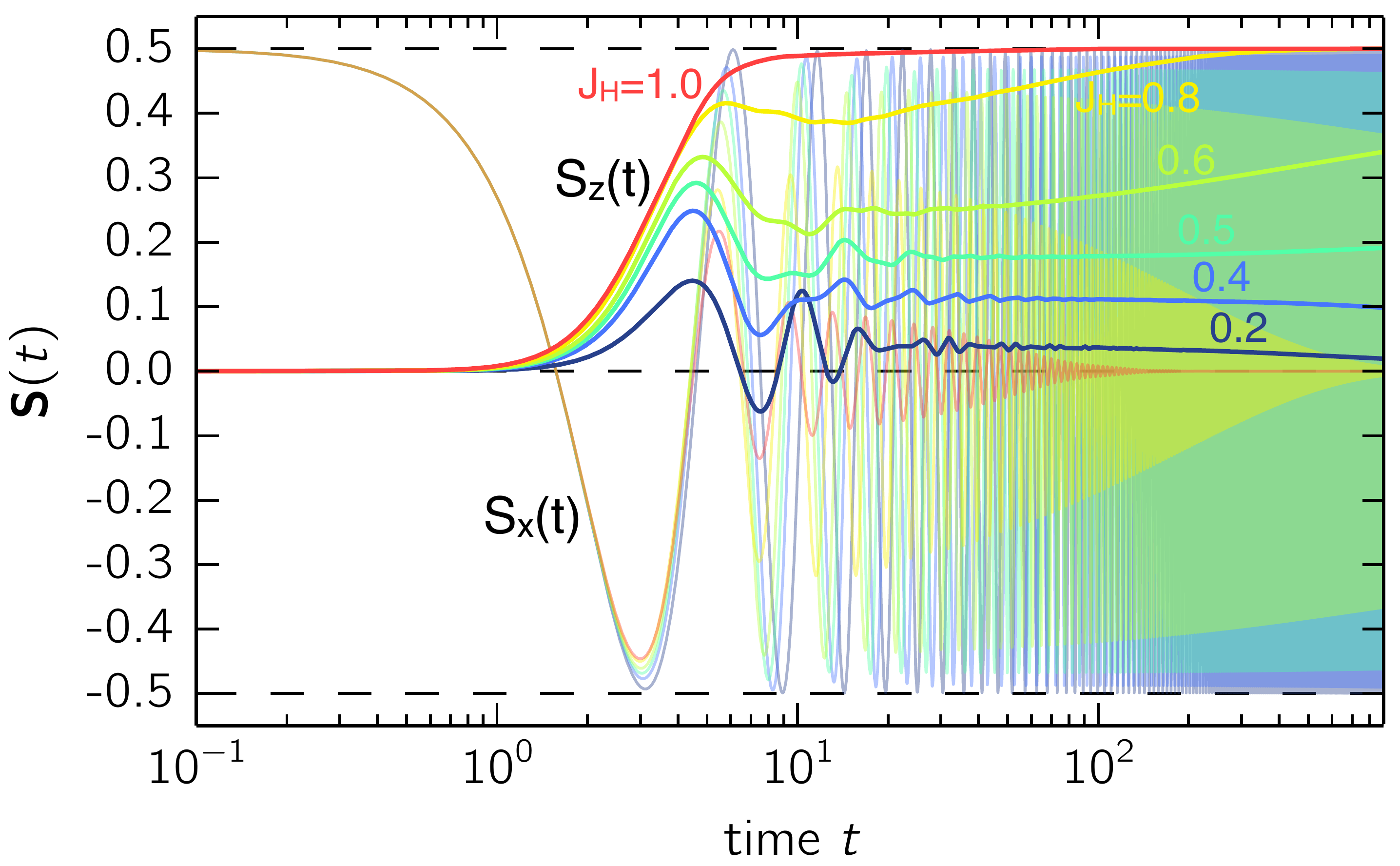}
\caption{
Classical spin dynamics for $J=1$ and different $J_{\rm H}$. 
} 
\label{fig:heisd}
\end{figure}

Fig.\ \ref{fig:heisd} displays the spin dynamics resulting from the full model
\begin{equation}
H 
= 
H_{\rm s} + H_{\rm s-spin} 
=
H_{\rm s} 
+
J \ff s_{i_0} \ff S 
-
\ff B \ff S \; ,
\label{eq:shei}
\end{equation}
as obtained by the linear-response approach.
One clearly notes that for $J\lesssim J_{\rm c} \sim 0.5$ (corresponding to $U_{\rm c} \sim 8$) the time dependence of $S_{z}$ develops a prethermalization-like plateau on an intermediate time scale $t \sim 100$ (in units of $1/J_{\rm H}$). 

For the Heisenberg model, using the scaling property of $\chi_{\rm loc}(t)$, it is easily possible to perform calculations up to $t = 1000$.
On this longer time scale, it is clearly visible (see Fig.\ \ref{fig:heisd}) that $S_{\rm z}$ does {\em not} approach a constant value asymptotically. 
For $J_{\rm H} = 0.4$ and $J_{\rm H} = 0.2$ the $z$-component of $\ff S(t)$ is even found to decrease and appears to approach the trivial solution $S_{z}(t) \equiv 0$. 

However, it is straightforwardly seen that a ``stationary state'' of the form
\be
\ff S(t) = S_{z} \hat{z} + S_{\perp} \cos(\omega t + \varphi) \hat{x} + S_{\perp} \sin(\omega t + \varphi) \hat{y}
\ee
with arbitrary parameters $S_{\perp}, \omega, \varphi$ and with constant (time-independent) $S_{z}$ does not solve the integro-differential equation (5) of the main text for $t \to \infty$. 
There is one exception only, namely the trivial case where $\chi_{\rm loc}(t) \equiv 0$ which can be realized, up to arbitrarily long times, in the limit $J_{\rm H} \to 0$. 

For small but finite $J_{\rm H}>0$, we therefore expect that the classical spin develops a dynamics on an extremely long time scale $t\gg 10^{3}$, the onset of which is already seen in Fig.\ \ref{fig:heisd}, which finally terminates in the fully relaxed state with $\ff S(t) \to \ff S_{0} \uparrow\uparrow \ff B$.

It is in fact easy to see from the integro-differential equation that, if there is spin relaxation to a time-independent constant, $\ff S(t) \to \ff S_{0}$ for $t\to \infty$, the relaxed state has $\ff S_{0} = 0.5 \hat{z}$. 
This implies that if there is complete relaxation at all, the spin relaxes to the equilibrium direction. 

\subsection{Oscillations at short times.}

As can be seen in Fig.\ \ref{fig:heisd} for weak $J_{\rm H}$, the $z$-component of the spin develops oscillations at short times, which can also be seen for the case of the Hubbard model (cf.\ Fig.\ 2 of the main text).
These oscillations can be understood in the following way: 
Inserting the expression (\ref{eq:short}) with $J'_{\rm H}=0$ for the behavior of $\chi_{\rm loc}(t)$ at short times into Eq.\ (5) of the main text, 
\begin{eqnarray}
\dot{\ff S}(t) = \ff S(t) \times \ff B - \frac23 J^{2} J_{\rm H} \langle \ff s_{i_{0}} \ff s_{i_{0}+1} \rangle \ff S(t)
\nonumber \\
\times
\int_{0}^{t} dt' (t-t') \ff S(t') 
+ \ca O(t^{3} J^{4}) \; ,
\end{eqnarray}
and approximating $\ff S(t)$ by the $J=0$ result $\ff S_{0}(t) = S (\cos\omega t,\sin \omega t,0)$ (with $\ff B = B \hat{z}$, $\omega=B$, $S=1/2$) in the second term on the right-hand side,
a straightforward calculations yields:
\begin{eqnarray}
  S_{z}(t) 
  &=& 
  \frac23 J^{2} J_{\rm H} \langle \ff s_{i_{0}} \ff s_{i_{0}+1} \rangle S^{2} \;
  \frac{
  \omega t \sin \omega t + 2 \cos\omega t - 2
  }
  {
  \omega^{3}
  }
  \nonumber \\
  &+ &
  \ca O(t^{4}J^{2}J_{H}^{2}) \; . 
\end{eqnarray}
This is found to perfectly describe the short-time oscillations for weak $J_{\rm H}$ in Fig.\ \ref{fig:heisd} and for strong $U$ in Fig.\ 2 in the main text. 
For longer times the oscillations are damped and eventually die out.

\end{document}